# Accelerated intermetallic phase amorphization in a Mg-based high-entropy alloy powder

Prince Sharma [a], Purvam Mehulkumar Gandhi [b], Kerri-Lee Chintersingh [b], Mirko Schoenitz [b], Edward L. Dreizin [b], Sz-Chian Liou [a], Ganesh Balasubramanian [a,1]

[a] *Institute for Functional Materials & Devices, Lehigh University, Bethlehem, PA 18015.*
[b] *Center of Materials for Advanced Energetics, New Jersey Institute of Technology, Newark, NJ 07102.*

**Abstract**

We describe a novel mechanism for the synthesis of a stable high-entropy alloy powder from an otherwise immiscible Mg-Ti rich metallic mixture by employing high-energy mechanical milling. The presented methodology expedites the synthesis of amorphous alloy powder by strategically injecting entropic disorder through the inclusion of multi-principal elements in the alloy composition. Predictions from first principles and materials theory corroborate the results from microscopic characterizations that reveal a transition of the amorphous phase from a precursor intermetallic structure. This transformation, characterized by the emergence of antisite disorder, lattice expansion, and the presence of nanograin boundaries, signifies a departure from the precursor intermetallic structure. Additionally, this phase transformation is accelerated by the presence of multiple principal elements that induce severe lattice distortion and a higher configurational entropy. The atomic size mismatch of the dissimilar elements present in the alloy produces a stable amorphous phase that resists reverting to an ordered lattice even on annealing.

---

[1] Corresponding author. 1 Farrington Square, Bethlehem, PA 18015. Email: biganesh@gmail.com

*Introduction:* Mg-Ti based alloys exhibit a promising blend of material properties relative to conventional titanium alloys, *viz.*, bioactivity, high strength, and reduced density, while possessing enhanced corrosion resistance relative to common magnesium alloys [1–7]. The atomic arrangement in amorphous Mg-Ti alloys lacks order, a characteristic of crystalline materials, resulting in exceptional structural strength [8,9] and producing lightweight components with notable durability [10,11]. The diminished presence of dislocations and grain boundaries improves the resistance to corrosion [8,9]. Likewise, amorphous Mg-Ti alloy powders possess high energy densities, making them suitable for applications like hydrogen storage and catalysis [12–16]. Consequently, Mg-Ti alloys hold immense potential for a wide array of applications ranging from orthopedic implants and aeronautical components to hydrogen storage vessels. Note that the synthesis of Mg-Ti alloys is nontrivial due to the positive enthalpy of mixing and limited solubility of magnesium in titanium, and vice versa [17,18]. To circumvent this challenge, nonequilibrium synthesis routes such as mechanical alloying, vapor quenching, physical layer deposition, and sputtering, are employed to fabricate Mg-Ti alloys [1,19–22]. Despite the successful synthesis, Mg-Ti alloys often remain in a metastable state post-synthesis and tend to undergo dissociation upon annealing [22,23].

Here, we present results from the synthesis of stable amorphous/nanocrystalline Mg-Ti alloy powders incorporating the principles underlying the formation of high-entropy materials. In particular, we leverage the concept of high configurational entropy that induces severe lattice distortion, sluggardizing the diffusion rates through the incorporation of different elements with significant concentrations into the lattice structure [24–30]. A Mg-Ti based high-entropy alloy powder, $Mg_{0.25}Ti_{0.25}Cu_{0.18}Zn_{0.18}Fe_{0.14}$ is synthesized using high-energy mechanical alloying. Characterization of the material reveals a structural transformation from a precursor intermetallic phase. This transformation involves the formation of defects, nanograin boundaries, and lattice expansion due to the presence of antisite disorder. Ground state density functional theory (DFT) calculations are employed to explain the mechanisms driving the synthesis and properties of the high-entropy alloy powder.

The novelty of our work lies in the strategic selection of alloy compositions to accelerate the synthesis of high-entropy amorphous alloy powder from an otherwise immiscible Mg-Ti rich metallic mixture using high-energy mechanical milling. The phase transformation from the precursor intermetallic structure to the amorphous phase is accelerated by the presence of multi-principal elements that induce severe lattice distortion and higher configurational entropy.

*Experimental:* High purity elemental powders (>99.95%, Sigma-Aldrich) of Mg, Ti, Fe, Cu and Zn are mixed in the predetermined ratio. The mixture is prepared in a shaker mill (SPEX SamplePrep 8000D) equipped with forced convective cooling for 2.5 (separate wet and dry runs), 5, 10, 15 and 20 hours. The powders are loaded in 65 mL hardened steel vials with 9.5 mm diameter hardened steel balls inside a glove box where oxygen concentration is maintained at 0.1%. The powder to ball mass ratio is 1:10 and 3 ml of n-hexane is used as the process control agent in wet runs. In dry runs, only dry starting commercial elemental powders and balls was loaded for milling for 2.5 hours. A representative schematic is shown in Figure 1. The 20 hour- milled powder is annealed at 500 $^O$C for 6 hours in argon atmosphere (MTI, GSL-1600X). The milled samples are then extracted within a glove box environment to mitigate any potential oxidation effects on the sample surfaces. The milled and annealed samples are characterized by X-ray diffraction (XRD, PANalytical; Cu-Kα target at λ = 1.54 nm), scanning electron microscopy (SEM, Hitachi 4300SE) equipped with energy dispersive X-Ray Spectroscopy (EDS, Octane elite), and transmission electron microscopy (TEM, JEOL 2100, operating voltage of 200 kV) to identify the phase and the composition of the synthesized powder.

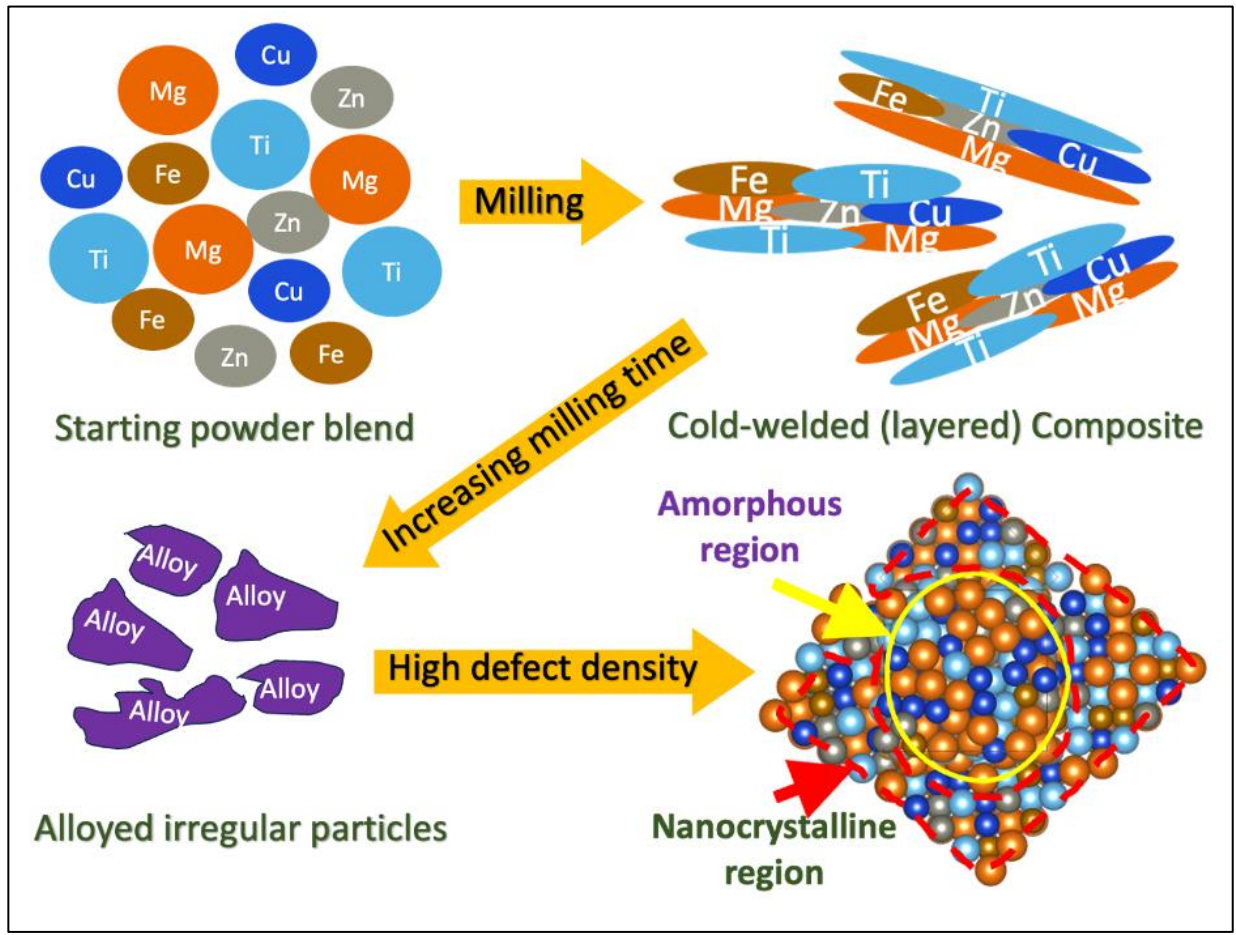

*Figure 1* Schematic of material evolution during high-energy ball milling. Continuous and highly energetic compressive impact on powder particles by the balls results in composite formation via cold welding. Further collisions lead to severe plastic deformation increasing the density of defects , dislocations, stacking faults, twins and grain boundaries, in the particles. An increasing fraction of such defects enforces an amorphous arrangement of the atoms. The atomic structure illustration highlights the amorphous region and the adjacent smaller nanocrystalline zones.

**Computational:** DFT calculations are performed with the Vienna ab-initio Simulation Package (VASP) [31,32]. Electronic configurations [Ne]$3s^2$, [Ar] $3d^2\, 4s^2$, [Ar] $3d^6\, 4s^2$, [Ar] $3d^{10}\, 4s^1$ and [Ar] $3d^{10}\, 4s^2$ are used to represent Mg, Ti, Fe, Cu and, Zn, respectively. The Projector Augmented Wavefunction (PAW) pseudopotential is employed to describe core electrons [33]. The Perdew, Burke, and Ernzerhof (PBE) generalized gradient approximation (GGA) exchange-correlation functional is utilized together with the Methfessel-Paxton smearing technique [34,35]. K-points mesh is generated using the Monkhorst-Pack approach with a precision level of 0.04 [36] to accurately integrate over the Brillouin zone. The self-consistent field procedure is adopted to solve the Kohn-Sham equations with a convergence criterion between two iterations established at $10^{-6}$ eV. The FCC, BCC and HCP unit cells are generated using the ATAT enabled Monte Carlo generator of Special Quasirandom Structures (mcsqs) [37]. We also construct an $AB_2$ type disordered hexagonal Laves phase unitcell with stoichiometry $(Mg_{0.5}Ti_{0.5})(Cu_{0.36}Zn_{0.36}Fe_{0.28})_2$ to examine the mechanism of mechanical alloying. For replicating an amorphous structure, a FCC supercell is heated to 5000 K (finite temperature DFT framework) and quenched to the ground state, followed by several steps of relaxation until the accuracy criterion is satisfied. Further, the formation energy per atom $\Delta E \frac{eV}{atom} = \frac{E(A_xB_yC_z) - xE(A) - yE(B) - zE(C)}{n}$ is derived from the differences of elemental ground state energies from that of the alloy structure and normalized by number of atoms. The Vesta program is used to calculate the XRD pattern of the mcsqs and amorphous supercells [38].

The supercells of the lattices, and their formation energies per atom are presented in Figures 2(a) and (b), respectively. The total formation energy per atom for the amorphous structure is the lowest relative to the intermetallic, BCC, FCC and HCP counterparts. The alloy powder is least stable in an FCC phase, while moderately stable as BCC and HCP crystals. Figure 2(c) exhibits the computed XRD profiles corresponding to the supercells described in Figure 2(a). The XRD pattern for the different crystal structures assumes characteristic peaks. In other words, the calculated XRD patterns of monocrystalline phases manifest themselves through the precisely defined peaks devoid of any distinct broadening. In contrast, for the amorphous counterpart, a notable lack of well-defined structural arrangement contributes to a diffusive scattering motif, replacing the sharp diffraction peaks characteristic of the crystalline materials. Interestingly, the XRD profile of the amorphous alloy exhibits a prominent hump, offering significant insights on the corresponding atomic ordering and interatomic spatial distribution. Crystalline supercells assume a repetitive atomic orientation causing a constructive interference at specific angles. On the other hand, the inherent disorder in amorphous structures induces nondirectional scattering tendencies, thereby yielding an uninterrupted continuum of scattered intensity. An intriguing observation is that the peaks emanating from the intermetallic phase predominantly align themselves proximal to the hump, evincing the amorphous structural configuration. Figure 2(d) displays the experimental XRD of the high-entropy powder, mechanically alloyed in a shaker mill over different duration. It is important to note that the XRD pattern of the as-synthesized powder reproduces a hump akin to that predicted from the calculated XRD of the amorphous phase in Figure 2(c). Given that the amorphization occurs at a very early phase of the milling process, it is extremely challenging to accurately determine the elemental order of alloying. We highlight that in Figure 2(d) XRD of 2.5 and 5 hours milled powder, the 2θ peaks at $40.8^O$ and $44.16^O$ prove formation of $TiFe_2$ hexagonal Laves phase. We observe that a 2θ peak at $41^O$ in powder becomes a main peak with longer milling time, this suggests presence of other hexagonal intermetallic namely, $MgZn_2$ and $TiZn_2$. This peak shift from $40.8^O$ to $41^O$ is evidence of formation of a disordered intermetallic from $TiFe_2$ via insertion of Mg and Zn **.** Figure 2(d) indicates the presence of a broad hump initiating at around 35º and continuing till 45º where it assumes sharp shoulder, implying some degree of crystallinity. For the sample characterized after 2.5 hours of dry milling, the hump arises even before 35º while the sharp peak at 45º produces reduced intensity; we conjecture this observation to be a result of powder caking *i.e.,* excessive cold welding of the powder on the vial, and hence subsequently synthesis is performed by wet milling only. After 5 hours of wet milling, the sharp peak is merged inside the hump suggesting an enhanced number of defects and a drastic reduction in crystallite size, transforming into the amorphous phase. The XRD of annealed sample exhibits a small shoulder in hump similar to 5-hours and 10-hours sample, suggesting no significant effect of temperature on amorphous alloy powders. We acknowledge that the XRD pattern obtained for the alloy powder resembles that for metallic glasses [39–43].

A narrow region of the computed XRD of the intermetallic phase is reported on the same set of axes with hump section measured the from experimental XRD in Figure 2(e). The characteristic peaks from the simulations, recorded around the amorphous region, concur with the results obtained from the synthesized samples, corroborating that the intermetallic phase is a precursor for the high-entropy alloy. We substantiate this insight with empirical thermodynamics. Figure 2(f) lists the pairwise enthalpies of mixing ($\Delta H_{mix}$) for the elements in the alloy based on Miedema's model [17]. $\Delta H_{mix}$ for all binary elemental combinations of Ti, except with Mg, are strongly negative suggesting a preferential formation of an intermetallic structure. On the other hand, Mg can form stable compounds only with Cu and Zn. Note that from the size of the elemental atoms presented as Figure 2(g) and inferences from XRD in Figure 2(d), the comparable atomic radii of Mg and Ti enable them to occupy the 'A' site of an 'AB$_2$' type intermetallic compound, whence Cu, Zn and Fe with similar atomic radii situate on the 'B' site of the lattice. Thus, an intermetallic structure of AB$_2$ Laves phase type, where A = {Mg, Ti}, B = {Fe, Cu, Zn}, *i.e.*, (Mg$_{0.5}$Ti$_{0.5}$)(Cu$_{0.36}$Zn$_{0.36}$Fe$_{0.28}$)$_2$ is formed prior to amorphization through continued milling. The theory is further validated by TEM characterization, as discussed below.

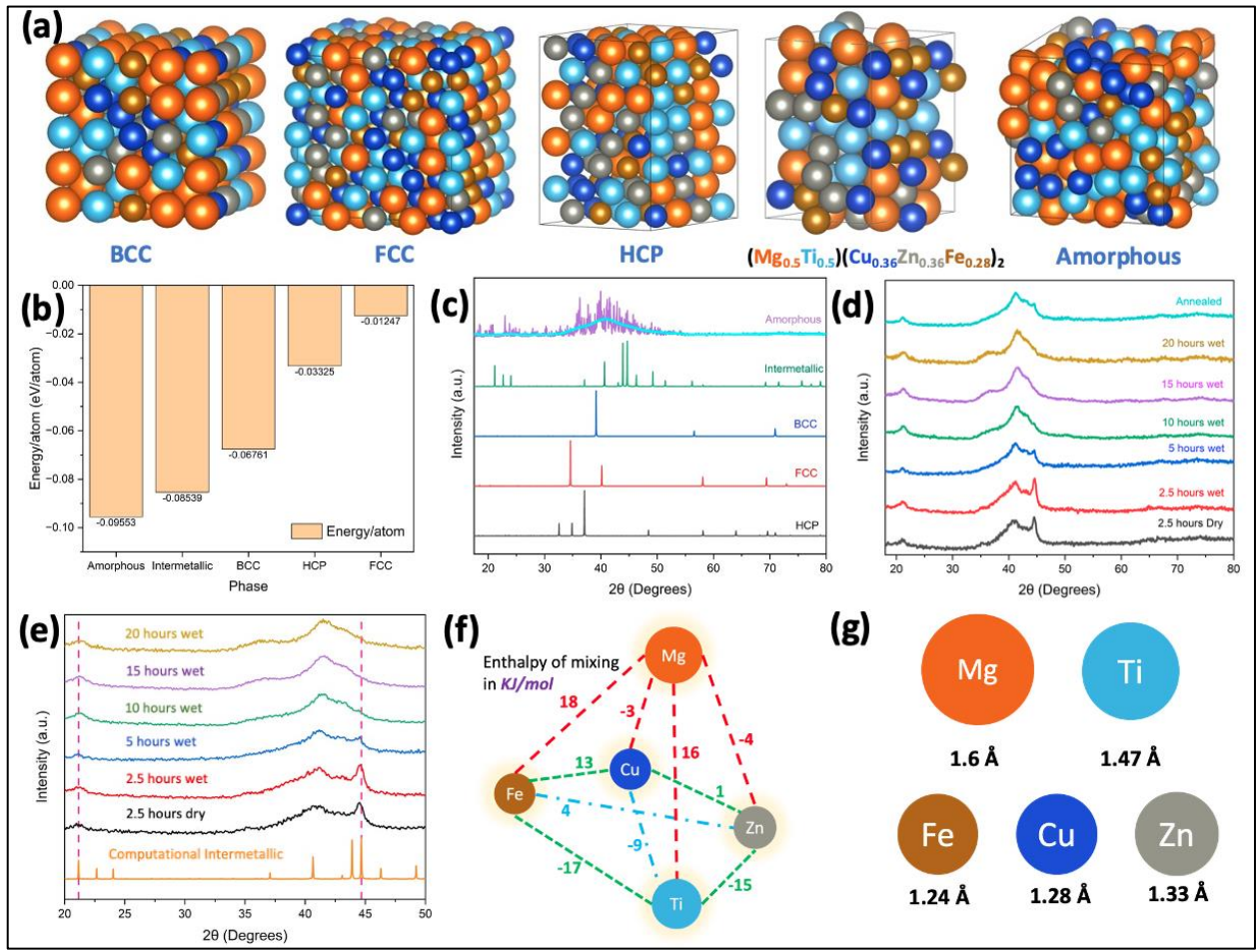

***Figure 2*** *The high-entropy alloy supercells, and the corresponding formation energies and computed X-ray diffraction patterns. (a) The BCC, FCC, HCP, intermetallic and amorphous phases replicated in the different supercell for the $Mg_{0.25}Ti_{0.25}Cu_{0.18}Zn_{0.18}Fe_{0.14}$ high-entropy alloy (orange for Mg, blue for Ti, indigo for Cu, grey for Zn and brown represents Fe). (b) The formation energy per atom for the five phases as predicted from first principles at the ground state. The amorphous and intermetallic phases are energetically favorable relative to the less stable crystalline structures. (c) The computed XRD of the five possible alloy phases superimposed with the XRD for the amorphous alloy. (d) XRD characterization of the ball-milled $Mg_{0.25}Ti_{0.25}Cu_{0.18}Zn_{0.18}Fe_{0.14}$ high-entropy alloy powder subjected to different milling times and annealed after milling. (e) A section of the experimental XRD presented in (d) is juxtaposed with the corresponding region of the computed XRD for the intermetallic structure. (f) Pairwise enthalpy of mixing $\Delta H_{mix}$ (kJ/mol) for*

*the constituting elements in the alloy. (g) Atomic radii of the principal elements in the high-entropy alloy.*

Figures 3 (a, b) present the SEM and EDS micrographs of the 2.5 and 20 hours wet milled powders, respectively. The powder morphology from the SEM images suggests that the particles are flaky and irregular in shape due to the continuous deformation they experience from the impact between the balls, and between the ball(s) and vial. The alloys assume a rather homogeneous elemental distribution, albeit the uneven heights of the various powder particles cast a contrast in the SEM images. We additionally perform TEM characterization of the milled powder. Figure 3(c) displays the TEM dark field (DF) image of the powder wet milled for 2.5 hours, and Figure 3(d) shows the corresponding selected area electron diffraction (SAED) pattern recorded from Figure 3(c). The SAED pattern reveals both the diffuse and broadened rings and some Bragg diffraction spots appeared to form a ring-like pattern. The first diffraction ring, as marked by the yellow bar, can be indexed as (112) plane of $TiFe_2$, which is consistent with the XRD result in Figure 2(d), indicating the formation of $TiFe_2$ hexagonal compound at a very initial stage of milling. Thus, we perform the TEM DF imaging, which selects one portion of diffraction spots to form imaging and reveal strong contrast with the particle size of 7~15 nm in contrast to the amorphous area generated less contrast due to strong disorder (and/or a high density of defects). The formation of $TiFe_2$ is also supported by a large negative enthalpy of mixing shown in Figure 2(f) as well as the XRD in Figure 2(d). The TEM images for the 20 hours wet milled powder, presented in Figures 3(e, f), reveal the presence of a large amorphous area (highlighted in yellow) fringed by crystalline phase (encircled by a red curve). The SAED patterns corresponding to the two marked regions, shown in Figures 3(g, h), confirms the existence of the amorphous phase and the presence of crystalline intermetallic compounds as identified as mixed of $TiZn_2$ and $MgZn_2$ as a kind of type $AB_2$ structure.

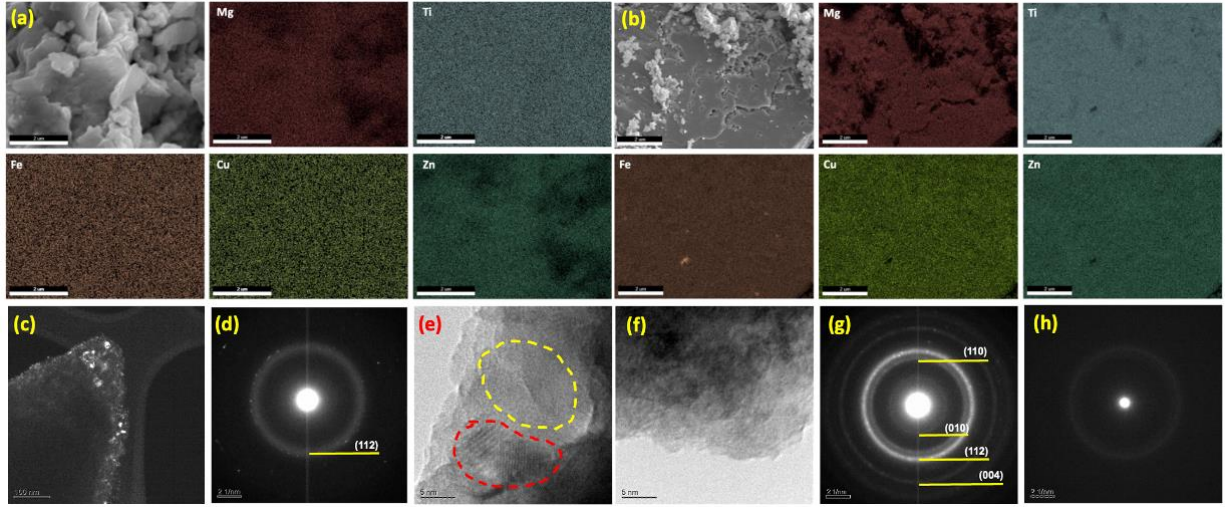

*Figure 3 SEM, EDS and TEM characterization of the mechanically alloyed $Mg_{0.25}Ti_{0.25}Cu_{0.18}Zn_{0.18}Fe_{0.14}$ high-entropy alloy powder. (a) SEM and EDS micrographs of 2.5 hours wet milled powder; (b) SEM and EDS micrographs of 20 hours wet milled powder; (c) TEM dark field images of 2.5 hours wet milled powder. The sharp contrast in (c) highlights the crystallinity region due to dark-field imaging; (d) SAED pattern confirming formation of $TiFe_2$ hexagonal phase; (e and f) TEM images of the 20 hours milled powder reproducing amorphous regions highlighted with a yellow curve, and crystalline phase on the periphery represented by the red curve; (g and h) SAED pattern for nanocrystalline and amorphous sections, respectively. The SAED pattern of the nanocrystalline can be indexed as the hexagonal, $TiZn_2$ and $MgZn_2$ intermetallic compounds, proving the transformation of an initial intermetallic phase by amorphization.*

The process of ball milling is accompanied by the introduction of defects in the lattice of the powder particles [44,45]. A crystalline defect is essentially a discontinuity in the ordered atomic configuration of a solid such that at the site of the defect, the symmetry of the crystal disrupts. Additionally, the alloy powder experiences complementary processes such as particle flattening, cold welding, plastic deformation, and fracturing [44–48]. Herein, the movement of atoms between the welded layers, known as interdiffusion, dominates the elemental self-diffusion, and is attributed to the profound impact energy and very high shear generated at the microscale during the milling process. Also, the significantly negative heat of mixing (Fig 2f) of the elements augments the interdiffusion, facilitating the formation of either an amorphous structure or a nonequilibrium crystalline phase. Ultimately, stability is driven by the energetic state, with the structure that exhibits the lowest free energy becoming the prevailing material phase.

The superimposition of the intermetallic peaks with the amorphous hump in Figure 2(e) suggests that the phase transition is of type III amorphization [44], where an intermetallic is transformed to an amorphous phase by antisite disorder and nanograin boundaries. Here, Ti forms stable intermetallic compounds with Fe, Cu and Zn. Although Mg has a relatively smaller affinity for mixing, it can form stable compounds with Cu and Zn. During the initial milling all the elements form a complex intermetallic with a highly distorted lattice, particularly since Fe, Cu and Zn do not form intermetallics amongst each other. During milling, the lattice distortion promotes antisite defects that are a prerequisite for amorphization [44,49,50].

The XRD in Figure 2(d) substantiates the above concept as there is an increase in the size of hump with increasing milling times, which is indicative of lattice expansion. Also, from first principles, we note that the FCC lattice has a unit cell length of 15.83 Å, while the amorphous structure derived from the former produces a unit cell of length 16.11 Å, i.e., lattice expansion because of antisite disorder causing amorphization. The energy transferred from the high energy impact is manifested towards the formation of antisite disorder and creating nanograins. We estimate the grain size from Figure 3(e) to be < 5 nm, implying that a large density of nanograin boundaries is formed by the mechanical alloying. Note that most intermetallic/alloy powders that are transformed into an amorphous phase via mechanical milling, can be reverted to their crystalline form by annealing [22,51–56]. In this instance, the influence of annealing on the amorphous structure of the alloy powder is scarcely noticeable, as indicated by the XRD analysis of the annealed sample in Figure 2(d). This is due to the inclusion of multiple principal elements (Fe, Cu and Zn) in high-entropy alloy, which contributes to the atomic mismatch, resulting in a very stable amorphous phase.

***Conclusions:*** In summary, the role of high-energy mechanical milling in promoting intermetallic amorphization for a stable nanocrystalline/amorphous high-entropy alloy is examined through a combination of first principles, materials theory and experiments. While amorphization of an intermetallic requires longer milling times, we exploit the high configurational entropy due to multiple elements in the alloy as well as their stronger affinity towards one solvent (Ti) relative to the other (Mg), to process an accelerated amorphization for a stable amorphous alloy. The transition from the precursor intermetallic structure to the amorphous phase is driven by antisite disorder and nanograin boundaries. The stability of the amorphous state and the atomic size mismatch due to the presence of diverse elements, renders the alloy resistant to a reversal to a crystalline lattice upon annealing.

**ACKNOWLEDGEMENTS**

The work was supported in part by the National Science Foundation (NSF) award # CMMI-1944040. We thank the Texas Advanced Computing Center (TACC) for providing Frontera resources that have enabled the computational results reported here.

**DATA AVAILABILITY**

The authors will make available, upon request, the data used in the applications described in this work. It is understood that the data provided will not be employed for commercial use.

**CODE AVAILABILITY**

The authors will make available, upon request, the code used in the applications described in this work. It is understood that the code will not be employed for commercial use.

**COMPETING INTERESTS**

The authors declare no Competing Financial or Non-Financial Interests.